\documentclass[letterpaper,12pt]{extarticle}  

\usepackage{import}
\usepackage{local}
\usepackage[left=.75in,top=.75in,bottom=.75in,right=.75in]{geometry}

\usepackage[utf8]{inputenc}
\usepackage[T1]{fontenc}

\usepackage{float}
\usepackage{physics}
\usepackage{graphicx}
\usepackage{dcolumn}
\usepackage{bm}
\usepackage{mathrsfs}
\usepackage{amssymb}
\usepackage{amsmath}
\usepackage{mathpazo}
\usepackage[normalem]{ulem}
\usepackage{wrapfig}

\usepackage[compress,square,sort,comma,numbers]{natbib}

\usepackage{amsmath}
\usepackage{mathtools}          
\usepackage{mathpazo}           
\usepackage[separate-uncertainty,detect-family=true,mode=math]{siunitx}
\usepackage{bm}
\renewcommand{\vec}[1]{\bm{#1}}

\usepackage{hyperref}
\usepackage{bookmark}
\usepackage{lipsum}

\usepackage{cleveref} 
\crefname{equation}{Eqn.}{Eq.}
\Crefname{equation}{Equation}{Equations}
\crefname{figure}{Fig.}{Figs.}
\Crefname{figure}{Figure}{Figures.}
\renewcommand{\vec}[1]{\bm{#1}}

\hypersetup{%
  breaklinks=true,
  colorlinks=true,
  linkcolor=black,
  filecolor=black,
  urlcolor=black,
  citecolor=black,
  pdfborder=0 0 0,
}

\begin{document}

\title{\LARGE Measuring response functions of active materials from data}
\author{%
\textbf{
Mehdi Molaei\textcolor{Accent}{\textsuperscript{*,1,2,3,**}}, %
Steven A. Redford \textcolor{Accent}{\textsuperscript{*,2,3,4,**}},
Wen-hung Chou \textcolor{Accent}{\textsuperscript{2,3,4}}
Danielle Scheff \textcolor{Accent}{\textsuperscript{2,3,5}}
Juan J. de Pablo \textcolor{Accent}{\textsuperscript{1}}
Patrick W. Oakes \textcolor{Accent}{\textsuperscript{6}}
Margaret L. Gardel \textcolor{Accent}{\textsuperscript{1,2,3,5,**}}
}\\
\begin{small}
\textcolor{Accent}{\textsuperscript{1}}Pritzker School of Molecular Engineering, University of Chicago, Chicago, IL 60637, U.S.A. \\
\textcolor{Accent}{\textsuperscript{2}}James Franck Institute, University of Chicago, Chicago, IL 60637, U.S.A. \\ 
\textcolor{Accent}{\textsuperscript{3}}Institute for Biophysical Dynamics, University of Chicago, Chicago, IL 60637, U.S.A. \\
\textcolor{Accent}{\textsuperscript{4}}Graduate Program in Biophysical Sciences, University of Chicago, Chicago, IL 60637, U.S.A. \\
\textcolor{Accent}{\textsuperscript{5}}Department of Physics, University of Chicago, Chicago, IL 60637, U.S.A. \\
\textcolor{Accent}{\textsuperscript{6}}Department of Cell and Molecular Physiology, Stritch School of Medicine, Loyola University Chicago, Maywood, IL 60153, U.S.A. \\
\textcolor{Accent}{\textsuperscript{*}} These authors contributed equally \\ 
\textcolor{Accent}{\textsuperscript{**}}Correspondence: \textcolor{Accent}{mehdi.molaei@gmail.com, redford@uchicago.edu, gardel@uchicago.edu} \\ \end{small}
}

\maketitle

\begin{onehalfspacing}

\section{abstract}
From flocks of birds to biomolecular assemblies, systems in which many individual components independently consume energy to perform mechanical work exhibit a wide array of striking behaviors. Methods to quantify the dynamics of these so called active systems generally aim to extract important length or time scales from experimental fields. Because such methods focus on extracting scalar values, they do not wring maximal information from experimental data. We introduce a method to overcome these limitations. We extend the framework of correlation functions by taking into account the internal headings of displacement fields. The functions we construct represent the material response to specific types of active perturbation within the system. Utilizing these response functions we query the material response of disparate active systems composed of actin filaments and myosin motors, from model fluids to living cells. We show we can extract critical length scales from the turbulent flows of an active nematic, anticipate contractility in an active gel, distinguish viscous from viscoelastic dissipation, and even differentiate modes of contractility in living cells. These examples underscore the vast utility of this method which measures response functions from experimental observations of complex active systems.

\newpage

\section{Introduction}

Active systems can be found at various scales across the natural world. From flocks of birds and schools of fish \cite{cavagna_bird_2014,marchetti_hydrodynamics_2013}, to swarms of bacteria \cite{dombrowski_self-concentration_2004, wensink_meso-scale_2012} and the protein filaments inside of cells \cite{needleman_active_2017}, a system is considered active if individual agents-- birds or molecular motors-- locally convert energy into mechanical work \cite{ramaswamy_mechanics_2010}. These local injections of energy produce forces, flows, and dynamic patterns on scales much larger than the active components themselves \cite{giomi_geometry_2015}. While all active systems share an underlying character, the structure and dynamics exhibited by each is a consequence of their specific mechanisms of energy injection, long range structural interactions, hydrodynamic milieu, and boundary conditions \cite{marchetti_hydrodynamics_2013}. Understanding how these factors conspire to produce the emergent phenomena we observe is an important challenge not just for understanding complex natural systems but also for designing novel and potentially autonomous materials. The myriad length and time scales often present, however, make characterizing the mechanics and dynamics of active systems difficult. 

When trying to gain access to length and time scales of active systems, the most easily accessible and plentiful data available are dynamic fields (e.g. velocity or intensity fields). Many methods have been developed to analyze these fields \cite{crocker_two-point_2000,cerbino_differential_2008,kadanoff_hydrodynamic_1963}. Correlation functions in particular have proven useful for extracting both length and time scales from velocity fields \cite{kadanoff_hydrodynamic_1963-1}. In the process, however, these methods reduce the information in complicated experimental fields to a few specific dimensions, limiting the insights we can glean. A complete picture of active dynamics would report the localized response of the system to an active perturbation. In thermally driven systems, material responses can be obtained from microrheology \cite{zia_active_2018,mason_linear_1995,gittes_microscopic_1997}. Two-point microrheology enables measurement of length and time scale dependent responses however the paucity of data necessitates spatial averaging over radial dimensions \cite{crocker_two-point_2000}. Recently, correlated displacement velocitometry was introduced which overcomes this limitation by considering the motion of many particles with respect to an averaged source \cite{molaei_interfacial_2021}. This method allowed for the measurement of interfacial flow around thermally driven colloids in two dimensions. By taking into account the internal directions of a displacement field this method effectively extended correlation analysis to measure the response function of a simple fluid near thermodynamic equilibrium. 

Here we extend this method to measure the response of active systems with unknown material properties far from equilibrium. We show that the idea introduced in ref. \cite{molaei_interfacial_2021} can be applied to any number of fields readily obtained from experimental data. Furthermore, by considering cross correlations we measure material response not only to displacement but other salient perturbations. We utilize this method to elucidate material and dynamical responses in a range of active materials composed of the biopolymer F-actin driven by myosin molecular motors. Specifically, in an active extensile nematic we extract the time-dependent length scales associated with vortical and shear perturbation that provide insight into turbulent flows. In a contractile active gel, we find that the ratio of these two critical length scales presage the onset of contractility. Furthermore, considering the temporal decay of these response functions allows us to distinguish viscous from viscoelastic dissipation in the nematic and the gel, respectively. Finally we show that this method can be harnessed to probe the mechanics of the actomyosin cytoskeleton in living cells. While we focused our attention here on active actomyosin materials this method should be of broad utility in the study of complex systems. 


\section{Method and Results}
\subsection{Measurement of response functions in 2D active materials}

\begin{figure}[bt!]
    \centering
    \includegraphics[width=0.75\textwidth]{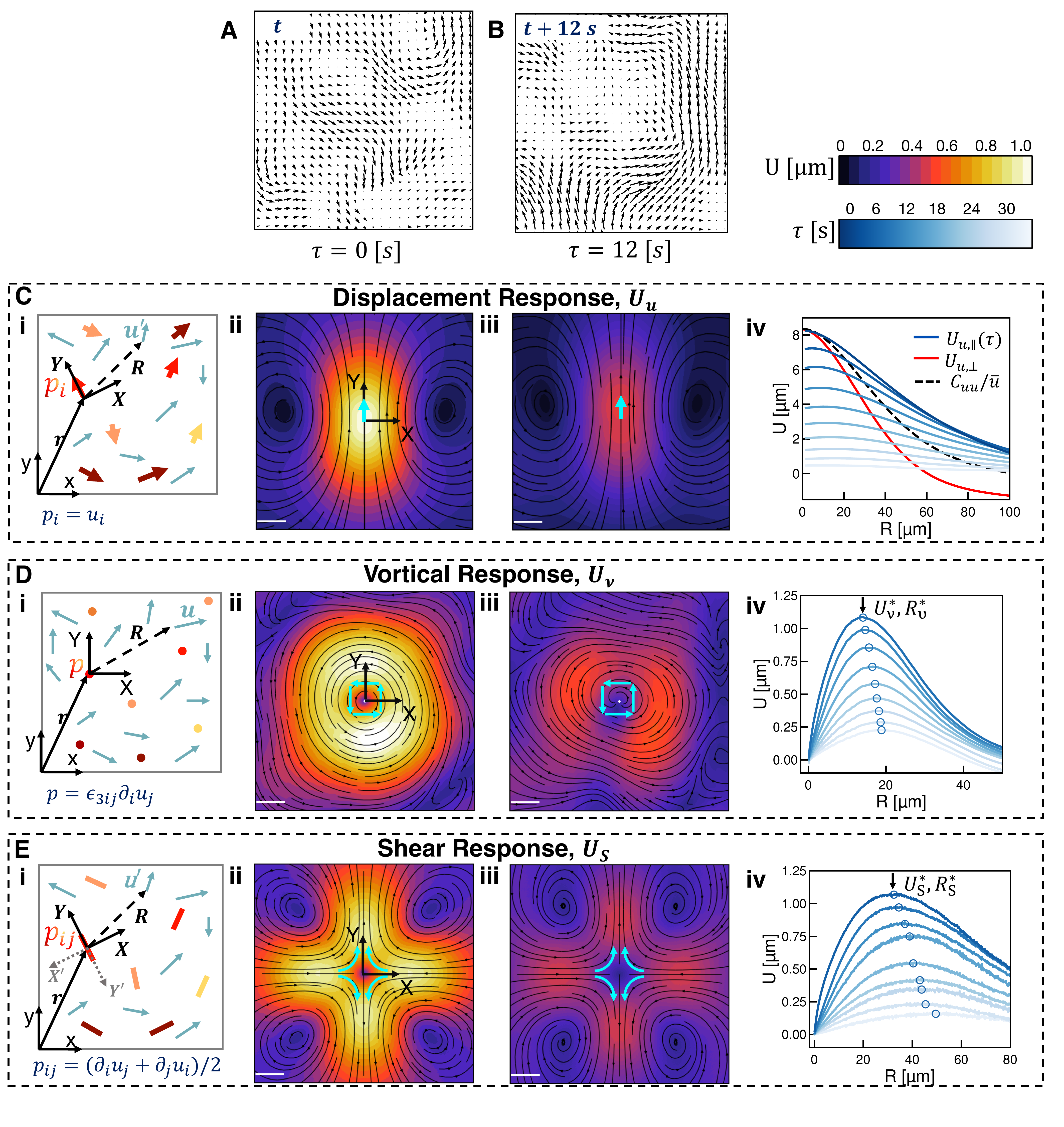}
    \caption{\textbf{Measuring directionally rectified correlations reveals response functions in active materials.} 
    A,B) Displacement fields measured by optical flow from fluorescence microscope images of an active nematic liquid crystal at \(t = 0s\) (A) and \(t = 12s\) (B). Displacements measured over \(\Delta t = \SI{2}{s}\).
    (i) Schematic of coordinate transformation and ensemble formation for the response of the two dimensional displacement field \(\vec{u}\) to various perturbation fields \(\vec{p}\). \(\vec{p}\) is either the displacement field itself, \(u\) (C), the vorticity field, \(\nu\) (D), or the anisotropic strain rate field, \(\vec{S}\) (E). (ii) Equal time two dimensional response of the displacement field in (A) to each of the respective perturbations. Ensembles are constructed such that the \(Y\) axis in (i) is the same as (ii). Streamlines indicate the direction of the resulting response function and color indicates the magnitude; scale bars are \(\SI{5}{\micro\meter}\). (iii) The \(\tau = 12s\) time delayed response function. The perturbation coordinate system is set at \(\tau = 0s\) and the response is measured at \(\tau = 12s\).  (iv) One dimensional profiles of the responses calculated at various lag times. Lighter colors indicate longer lag times. C,iv) One dimensional profile is constructed by tracing along the major axis (blue). Note that the azimuthal average is simply the average of the major axis trace and the minor axis trace (red). D,iv) One dimensional profile is constructed by azimuthally averaging. Model free characteristic deformation \(U^*\) and length \(R^*\) scales at the various lag times indicated by open circles. E,iv) One dimensional profile constructed as a trace along the \(Y\) axis. The origin symmetry of the field shown is unique to divergence free systems. Note the large characteristic length scale in comparison to vortical deformation fields.
    }
    \label{figure1}
\end{figure}

To begin, we consider a displacement field for a two dimensional active material. \cref{figure1}A shows one such field from an experimental active system. 
 As typical active materials contain a number of length and time scales, our approach will be to extract as much information as we can from this important dynamic field. To do so, we will approach correlation analysis not as conventional autocorrelation but as generalized cross correlation between some field \(\vec{p}\) and the displacement field \(\vec{u}\). The real trick here will be to take special notice of any directional information in \(\vec{p}\). This directional information will help the correlation to represent an averaged response of \(\vec{u}\) to the specific perturbation represented by \(\vec{p}\). 

Because \(\vec{p}\) could in principle be a high rank tensor field and interpreting high ranked correlation functions is difficult (See SI), we introduce the general correlation function \(\vec{\chi}_{\vec{p}}\). 
 \begin{equation}
    \vec{\chi}_{\vec{p}}(\vec{R},\tau)=\langle p(\vec{r}_1,t) \vec{u}(\vec{r}_2,t+\tau) \delta(\vec{R}-\vec{r}'_{12})\rangle_{\vec{r}_1,\vec{r}_2,t},
    \label{mod_corr}
\end{equation}
here \(p\) is the scalar magnitude of the field \(\vec{p}\) which renders \(\vec{\chi}_{\vec{p}}\) the same rank as \(\vec{u}\). \(\langle \cdot \rangle_{\vec{r}_1,\vec{r}_2,t}\) denotes averaging over space and time. Since \(p\) and \(\vec{u}\) are often discrete measurements, the average is computed by binning our data over a window of chosen size, which is represented by the finite width delta function \(\delta\). \(\vec{R}\) is the location in a new Euclidean space with the same dimensionality as \(\vec{u}\). Finally, \(\vec{r}'_{12}\) is the distance vector between \(\vec{p}\) and \(\vec{u}\), which will be constructed in this new coordinate system to report on the location of \(\vec{u}(\vec{r}_2)\) with respect to the position and heading of \(\vec{p}(\vec{r}_1)\). The purpose of \(\vec{r}'_{12}\) is to center the average such that each \(\vec{p}(\vec{r_1})\) is located at the origin of \(\vec{R}\). In practice, we will be dealing not with \(\vec{\chi}\) itself but rather the normalized correlation field \(\vec{U}_{\vec{p}}=\vec{\chi}_{\vec{p}}/\sqrt{\langle p^2\rangle}\). In tandem with the field \(\vec{p}\) the choice of coordinate system in which \(\vec{r}'_{12}\) is measured is a critical one and has important ramifications for how we interpret these correlations. To see why let us consider a concrete example. 

We can consider the displacement-displacement auto-correlation function in this new setting; \(\vec{U}_{\vec{u}}=\vec{\chi}_{\vec{u}}/\sqrt{u^2}\). Our objective is to understand how the displacement field responds to each specific \(\vec{u}(\vec{r_1})\). This means we want to take into account not only the location \(\vec{r_1}\) but also the heading of \(\vec{u}(\vec{r_1})\). To do so, we set up a new coordinate system whose Y axis aligns with \(\vec{u}(\vec{r_1})\). That is, we define \(\vec{r_{12}}' = \vec{M} (\vec{r_2} - \vec{r_1})\), where \(\vec{M} = \begin{pmatrix} \cos(\theta_1) & -\sin(\theta_1)\\ \sin(\theta_1) & \cos(\theta_1) \end{pmatrix}\) is the rotation matrix in which \(\theta_1\) is the angle between \(\vec{u}(\vec{r_1})\) and the \(Y\) axis of the lab frame. By defining \(\vec{r}'_{12}\) in this way, \(\vec{U}_{\vec{u}}\) reports on the correlation of the displacement field with a perturbing vector pointing along the \(Y\) axis. By normalizing by \(\sqrt{\langle u^2\rangle}\) this correlation reports on the average behavior of the displacement field with respect to an impulse of defined direction and magnitude, (\cref{figure1}C). Put differently, the function \(\vec{U}_{u}\) reports on the averaged \textit{response} of the displacement field \(\vec{u}\) to a perturbation of unit magnitude along the \(Y\) axis. This procedure is closely related to what was utilized recently to measure the Stokeslet flow induced by Brownian motion of passive colloids at an interface \cite{molaei_interfacial_2021}. By taking this directional information into account, the resulting response function is not radially symmetric as one would expect from traditional auto-correlation. Rather, the resulting displacement response function is only reflectively symmetric about the \(Y\) axis (\cref{figure1}C,ii). As one might expect, the response to a displacement along one axis decays at different rates for different angles with respect to that impulse.
 The response in the longitudinal direction \(U_{u,\parallel}=U|_{X=0}\) propagates over the largest distance and in the transverse direction, \(U_{u,\perp}=U|_{Y=0}\) over the shortest distance (\cref{figure1}C,ii). 
The difference between these two length scales is mainly related to the hydrodynamic coupling of the active material to the viscous bulk fluids \cite{levine_dynamics_2002, chisholm_driven_2021, martinez-prat_scaling_2021}.
The ratio of these two length scales quantifies the ratio of the kinetic energy that is dissipated within the 2d system (active nematic in this case) and kinetic energy dissipated externally. 
These two spatial length scales, however, get convoluted when calculating the conventional displacement auto-correlation function \(C_{\vec{u}\cdot\vec{u}}=\langle \vec{u}(\vec{r})\cdot\vec{u}(\vec{r+R}) \rangle_r/\sqrt{\langle \vec{u}(\vec{r})\cdot\vec{u}(\vec{r}) \rangle}_r\) as all perturbations \(\vec{u}\) are treated identically regardless of heading. 
In fact, this traditional auto-correlation function is equal to the azimuthal average of the response function we consider, \(C_{\vec{u}\cdot\vec{u}}=1/2\pi\int_0^{2\pi}{U_Y}(\vec{R})d\theta=(U_{\perp}+U_{\parallel})/2\); where \(\theta\) is the angle in polar coordinates.  
By taking into account the headings of each element of the displacement field as we construct the correlation function we are able to access important two dimensional features of flow responses. 
Unfortunately, while this description of the auto-correlation response contains more information than the previous version, determining a length scale from these functions still requires a model. This stems from the fact that any auto-correlation function must inherently be maximally correlated with itself at the origin. 
Without any universal scaling describing the decay of this correlation in active systems we are left to approximate. We will show that such ambiguity may not exist in responses to higher order perturbations. 

To capture the response of the displacement field to high order modes of deformation, we explore the response of the displacement field to its own gradients, \(\nabla \vec{u}\). 
In a linear system, we begin by decomposing the displacement gradient tensor \(\partial_i u_j\) into the anisotropic symmetric traceless strain rate tensor, \({S}_{ij}=(\partial_iu_j+\partial_ju_i)/2-\partial_ku_k I_{ij}/2\), the isotropic symmetric strain rate tensor \({D}_{ij}=\partial_ku_k I_{ij}/2\), and the circulation tensor \({\Omega}_{ij}=(\partial_iu_j-\partial_ju_i)/2\). 
Here, \(\vec{I}\) is an identity tensor with the same rank as \(\partial_i u_j\). This choice of decomposition has great physical utility as in two dimensions it corresponds to separating the contributions of pure shear, normal, and vortical deformation from one another. 
Having performed this decomposition, we construct a family of response functions \(\vec{U}_{\vec{S}}, \vec{U}_{\vec{D}}, \vec{U}_{\vec{\Omega}}\) corresponding to the response of the deformation field to the different perturbations, each of which quantifies different flow structures in \(\vec{u}\). We will defer the physical interpretation of these response functions in the various systems for the later sections but will focus here on the methodology for constructing each. Specifically, let us focus on the critical question of how to choose a coordinate system. 

Among these tensors, \(\vec{D}\) -- which quantifies bulk contraction or extension --  does not have a unique eigendirection and can be fully described by its first principal invariant. Therefore it can be treated as a scalar field, \(D\). As scalar fields do not have headings, we cannot choose a meaningful angle about which to rotate. In such cases, we default to a translation of the lab frame (\cref{figure1}D,i). More precisely, in the case of a scalar perturbation field we simply take \(\vec{r_{12}}' = (\vec{r_2} - \vec{r_1})\).
A similar situation arises with the circulation tensor \(\vec{\Omega}\) in two dimensions. This tensor has only one pseudovector which is normal to the plane of observation with rotation rate equal to vorticity  \(\nu=\hat{\vec{e}}_z\cdot\vec{\nabla}\times \vec{u}\). 
As the pseudovector of \(\vec{\Omega}\) provides no extra information, we will use the scalar vorticity field, \(\nu\), throughout this work and choose the simple translation of the lab frame as our coordinate system (\cref{figure1}D,i).
In the case of both \(D\) and \(\nu\), the resulting response function is radially symmetric which is a product of the lab frame coordinate that we have chosen (\cref{figure1}D,ii). 
Intuitively, we expect a system's response to vortical perturbation, \(\vec{U}_{\nu}\) or normal deformation to be radially symmetric which is the case in the measured responses (\cref{figure1}D,ii).

Unlike its cousins, \(\vec{S}\) is not a scalar field and thus requires more care when choosing coordinates. \(\vec{S}\) possesses two eigenvectors which identify pure shear directions with eigenvalues \(\pm \lambda\) indicating shear rate.  
To construct \(\vec{\chi}_{S}\) in this case we setup a coordinate system similar to the case of displacement auto-correlation where the \(Y\) axis is aligned with the eigendirection corresponding to the positive eigenvalue, (\cref{figure1}E,i). It is important to note here that it does not truly matter which eigenvector we choose so long as we are consistent while constructing the response function. As a convention, we have chosen the eigenvector associated with positive shear as we will eventually turn this method on an extensile system.
The eigenvalue of \(\vec{S}\) in this direction, \(\lambda\), is used as the scalar field which normalizes the correlation. 
The shear response function we construct, \(\vec{U}_{\vec{S}}=\vec{\chi}_{\vec{S}}/\sqrt{\lambda^2}\) is therefore the response function of the system to a pure extensile shear perturbation at the origin pointing outwards along the \(Y\) axis as seen in \cref{figure1}E,i.
Since \(\vec{S}\) is a symmetric and traceless tensor, the resulting response function \(\vec{U}_{\vec{S}}\) is symmetric about both \(X\) and \(Y\) axes (\cref{figure1}E,ii). This is the high rank extension of what we have seen already. Scalar fields, having no internal directions in the plane, form radially symmetric correlation functions while vector fields yield only one axis of symmetry. This response to a second rank tensor which represents bidirectional motion yields a response function that is symmetric about two separate axes. 

So far we have only measured equal-time response functions. Evaluating the response after some lag time however is straightforward. 
Taking a perturbation field \(\vec{p}(\vec{r}_1)\) at time \(t\) the delayed response function is constructed from \(\vec{u}\) measured after some delay time \(t + \tau\).
A snapshot of the velocity field from \cref{figure1}A measured at \(\tau=\SI{12}{s}\) is shown in \cref{figure1}B. 
From these two displacement fields corresponding response functions \(\vec{U}_{\vec{u}}(\tau)\), \(\vec{U}_{\nu}(\tau)\), and \(\vec{U}_{\vec{S}}(\tau)\) are computed at \(\tau=\SI{12}{s}\) (\cref{figure1}C-E,iii). The spatiotemporal response functions as measured here provide useful information about spectral properties of the dynamics of active matter in the time domain, which has proven challenging so far to gain access to \cite{alert_active_2022}. 

Thus far then we have focused solely on technical implementation of this approach. We have demonstrated that by taking into account internal headings of a perturbing filed, we can use the familiar language of correlation functions to construct functions that report on the response of a displacement field to these specific perturbations. After decomposing the displacement gradient tensor into constituent parts we demonstrated how to construct similar response functions for various ranks of perturbing inputs. Finally, we have seen a simple procedure to extend this framework into the time domain. Having detailed these technical steps, we now utilize our analysis on real data. In each we will ask what information this new approach can reveal compared to previously established methods for a few well characterized active systems. 

\clearpage 
\subsection{Response Functions Reveal Characteristic Length Scales in Active Nematic Turbulence}
We begin by analyzing the flow structures of an active nematic liquid crystal, as their steady state dynamics are well characterized \cite{giomi_geometry_2015,marchetti_hydrodynamics_2013,zhang_interplay_2018}. Here we consider an extensile nematic composed of short actin filaments driven by mysoin motors \cite{kumar_tunable_2018,zhang_spatiotemporal_2021}. \Cref{figure1}A and B are snapshots of the displacement field \(\vec{u}\) of an active nematic measured over \(\Delta t=\SI{2}{s}\) at two different time points. We measure the displacement response \(\vec{U}_{\vec{u}}\) in this system and find a difference in decay length between the direction parallel and perpendicular to the averaged displacement vector (\cref{figure1}C,ii). The monotonic decay of these traces preclude the extraction of an unambiguous length scale as quantifying the decay would require a model or ansatz, \cref{figure1}C,ii. The non-monotonic shape of the vortical and shear responses overcome this limitation (\cref{figure1}D,E iv). 

The azimuthal average of the vortical response function \(U_\nu(R)\), exhibits a clear peak (\cref{figure1}D,iv). Since the helicity \(\langle \nu(\vec{r})\vec{u}(\vec{r})\rangle_{\vec{r}}\) in a 2D nematic field is zero, \(U_\nu\) starts from zero at the origin and rises to the characteristic magnitude \(U_{\nu}^*\) at the characteristic length scale \(R^*_{\nu}\), before decaying at large distances due to dissipation mechanisms \cite{alert_active_2022}. To put the characteristic length scale \(R^*_{\nu}=\SI{14.1}{\micro\meter}\) in context with established methods, we measure the distribution of vortex sizes using the standard Okubo-Weiss parameter and velocity winding number \cite{lemma_statistical_2019, giomi_geometry_2015, guillamat_taming_2017} (Fig. S1). Applying this method to our distribution we find an average vortex area of \(\SI{688\pm 9}{\micro\meter^2}\). This is in close agreement with the vortex area calculated from our model free vorticity length scale \(\pi (R^*_{\nu})^2=\SI{624}{\micro\meter^2}\). Thus the vortical response function reports a critical length scale that is commensurate with the radius of an average vortex in the system. 

The shear response \(\vec{U}_{\vec{S}}\) for an active nematic exhibits a high degree of reflective symmetry (\cref{figure1}E,ii). In this case \(\vec{U}_{\vec{S}}\) is symmetric about not only the \(X\) and \(Y\) axes as we would expect from the response to a second order tensor, but also the diagonal \(Y=X\). This symmetry arises from the incompressibility of the nematic film. Because \(\vec{U}_{\vec{S}}\) is diagonally symmetric, we consider only the one dimensional trace along the \(Y\) axis (\cref{figure1}E, iv). Similar to the vortical response, this one dimensional trace rises from zero at small length scales and decays due to various dissipation mechanisms giving a critical magnitude, \(U_{\vec{S}}^*\), and length scale, \(R_{\vec{S}}^{*}\), of flow structures associated with shear perturbation. In this case, this one dimensional response trace starts from zero at the origin because inertial advection \(\vec{u}\cdot\vec{\nabla}\vec{u}\) at small length scales is zero. It is interesting to note that in general \(R_{\vec{S}}^* > R_{\nu}^*\). This is because the shear deformation field \(\vec{U}_{\vec{S}}\) is coupled to pressure gradients in the system while the vortical deformation field, in the absence of inertial effects,  is not \cite{batchelor_introduction_2000, marchetti_hydrodynamics_2013}. 
Vortical perturbation at the origin only propagates by the curl of force density due to elastic, flow-alignment, and active stresses and thus is dictated in large part by the systems shear rheology \cite{martinez-prat_scaling_2021}. Because these scales are model-free and unambiguous they allow for comparisons between active nematics with different levels of activity and reveal a scaling law in nematic dynamics (Fig. S2) \cite{alert_universal_2020,koch_role_2021}.

\subsection{Critical Length Scales Anticipate the Onset of Contractility in an Active Gel}
\begin{figure}
    \centering
    \includegraphics[width=0.75\textwidth]{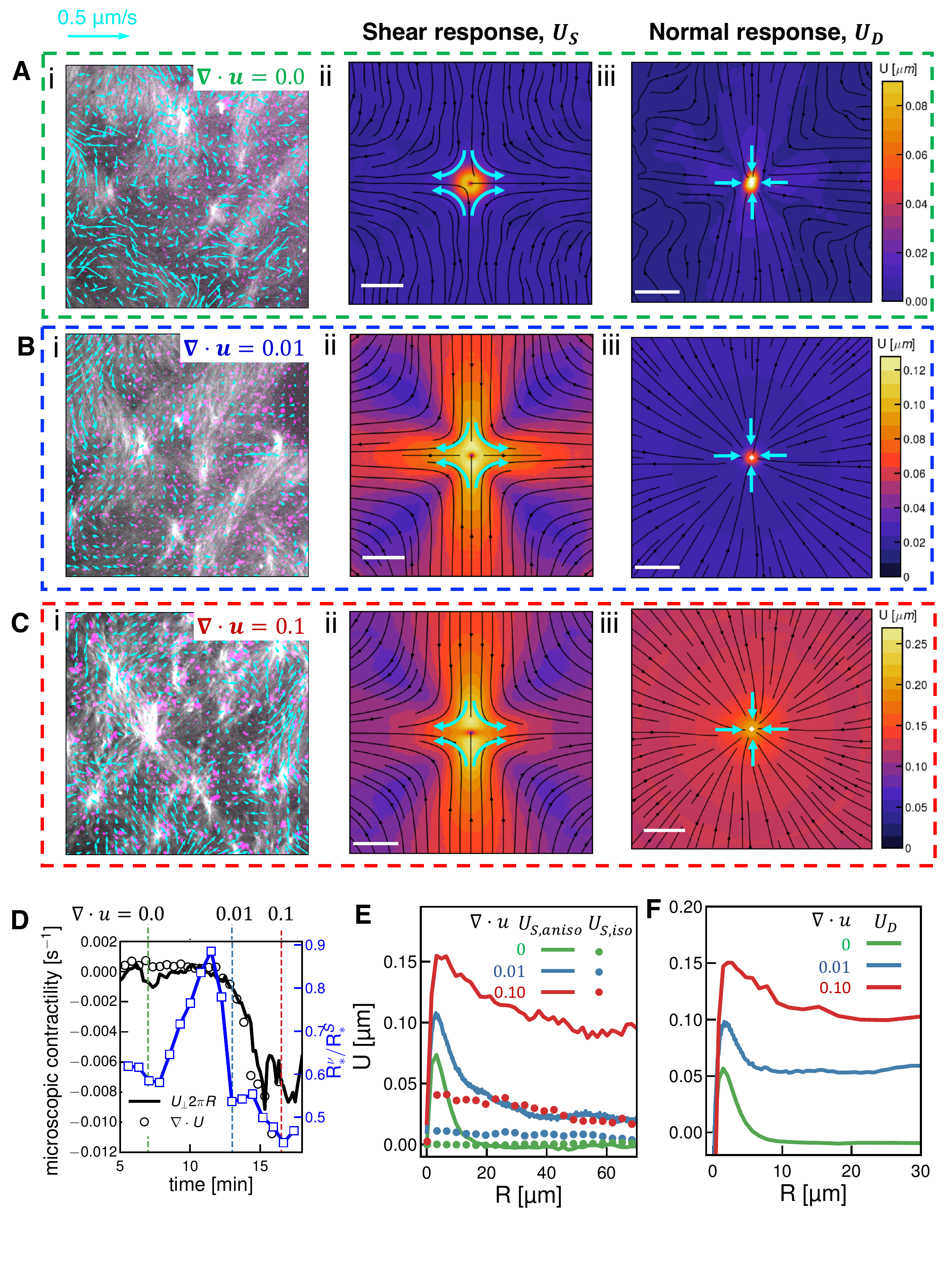}
    \caption{\textbf{Response functions identify key dynamic consequences of contraction in \textit{in vitro} actomyosin networks.} 
    i) Micrographs of fluorescent actin (gray) and myosin (magenta) overlaid with scaled velocity vectors (blue) for the active gel at various divergences (A-C). Colors correspond to divergences indicated in (D), each box is \(100\times100\SI{}{\micro\meter^2}\) of the field of view. ii) Equal time shear response for the velocity fields in (i). iii) Normal (compression) response for the velocity fields in (i). Streamlines indicate the direction of the resulting correlation field and color indicates the magnitude; scale bars are \(\SI{5}{\micro\meter}\). D) Divergence of the velocity field (black circles) as a function of time for a contractile active gel. The azimuthal average of the isotropic shear response per unit area at \(R=R^*_{\vec{S}}\), \(U_{\vec{S}, iso}(R^*_{\vec{S}})/R^*_{\vec{S}}\) (solid black line) agrees with the calculated divergence. The green, blue, and red dashed lines indicate the points taken as characteristic of the gel before contraction, at the onset of contraction, and deep in the contractile regime. Numerical labels indicate the value of divergence at these points. Ratio of characteristic length scales \(R^*_{\nu}/R^*_{\vec{S}}\) as a function of time (solid blue line, open blue squares). Time axis indicates elapsed time after the addition of myosin motors. (E) One dimensional traces of the symmetric (\(U_{\vec{S}, iso}\), solid lines), and anisotropic (\(U_{\vec{S}, aniso}\), dotted line) decomposition of the shear response function in (ii) as a function of distance \(R\). (F) One dimensional traces of the normal (compressional) response in (iii) as a function of distance \(R\).}
    \label{figure2}
\end{figure}

In contrast to active nematics that exhibit dynamics at steady state, the structure and dynamics of contractile active gels evolve over time (\cref{figure2} A-C,i, actin fluorescence in grey scale, myosin fluorescence in magenta, and Fig. S3) \cite{murrell_f-actin_2012,stam_filament_2017}. The gels we consider are composed of long actin filaments driven by Myosin II motors which crosslink and buckle filaments leading to bulk contraction \cite{murrell_f-actin_2012}. One interesting feature is that after the addition of myosin motors at \(t=\SI{0}{\minute}\), the gel does not contract immediately. Rather as myosin filaments settle onto the network and stresses slowly build up, the divergence of the displacement field \(\langle\vec{\nabla}\cdot \vec{u}\rangle\) remains negligible for a time before rapidly decreasing as the system irreversibly contracts (\cref{figure2}D, black circles). This delay has been a particular focus of studies on active gels, specifically the relation between the buildup and spectrum of internal stresses and the stability or contractility of the system \cite{seara_entropy_2018}. To investigate this transition in detail we measure the shear \(\vec{U}_{\vec{S}}\) and normal \(\vec{U}_D\) response at three representative stages of contractility where \(\langle\vec{\nabla}\cdot \vec{u}\rangle\) of  \(\sim 0\), \(0.01\), and \(0.1\) (\cref{figure2}D, vertical dashed lines). As we expect, the magnitude of the normal response \(\vec{U}_D\) increases as contractility builds reflecting the compression of the gel (\cref{figure2}A-C,iii; F). 


One striking feature of the contractile gel is that the shear response in the transverse direction decays more quickly than along the major axis of contraction (\cref{figure2}B,ii, comparing the \(X\) (transverse) and \(Y\) (contraction) axes). The asymmetry becomes even more pronounced for increasing divergence (\cref{figure2}B-C,ii). We can think about this deviation from symmetry as the addition of an isotropic response that represents the net contraction of the system due to shear perturbations. To isolate these different modes of deformation, \(\vec{U}_{\vec{S}}\) can be separated into an isotropic and anisotropic part \(\vec{U}_{\vec{S}}=[-U_{\vec{S} R, aniso}\cos{2\theta}+U_{\vec{S}, iso}]\hat{\vec{e}}_{R}+U_{{\vec{S}}_{\theta, aniso}} \sin{2\theta}\hat{\vec{e}}_{\theta} \), where \(\vec{U}_{\vec{S}, aniso}\) gives the strain dipolar modes, and \(\vec{U}_{\vec{S}, iso}\) gives the compressional mode. Splitting the response in this fashion allows us to track the progression of contraction in the gel directly from the shear response. Specifically, the azimuthal average of the isotropic shear response per unit area at \(R^{*}_{\vec{S}}\), \(U_{\vec{S}, iso}(R^*)/R^{*}_{\vec{S}}\), is equal to the average divergence, \(\langle\vec{\nabla}\cdot \vec{u}\rangle\), of the displacement field (\cref{figure2}D, solid black line). The isotropic shear response \(\vec{U}_{\vec{S}, iso}\) then, is reporting on the magnitude of contraction induced by local shear stress. Thus in this system, which is driven by the local sliding of pairs of filaments \cite{murrell_f-actin_2012}, this response captures the global divergence. It should be noted here the relationship between \(\vec{U}_{\vec{S}, iso}\) and \(\vec{U}_{D}\). The normal response \(\vec{U}_{D}\) measures the system's response to the contraction captured by \(\vec{U}_{\vec{S}, iso}\). Considering the azimuthal average of each, we find that both \(U_{\vec{S}, iso}\) (\cref{figure2}E, dots) and \(U_{D}\) (\cref{figure2}F, lines) increase as contractility increases. \(U_{\vec{S}, iso}\) decays weakly as a function of distance, reflecting the spatial dependence of local contractility (\cref{figure2}E,dots). \(U_{D}\) exhibits a clear peak indicating a length scale over which contractility drives maximal material deformation (\cref{figure2}F). A similar peak occurs in \(\vec{U}_{\vec{S}, aniso}\) which captures material rearrangements due to shear (\cref{figure2}E, solid lines). Furthermore, as contractile flows in the material increase, the energy scale and extent of this local shear response -- both \(R^*\) and \(U^*\) -- grow monotonically (Fig. S4). Interestingly, after these peaks none of these responses decay fully to zero (\cref{figure2}E,F). As noted in the nematic, at steady state the far field response decays to zero. The lack of such decay in the far field (\(R > R^*\)) indicates that the contractile gel dynamics are not at steady state.

In addition to specifically enumerating these mesoscale deformations due to contractile stress, the resolution of our measurement reveals features that seem to presage the onset of contractility. Recall that the length scale associated with shear response \(R^*_{\vec{S}}\) is typically larger than the length scale associated with the vortical response \(R^*_{\vec{S}}\) (\cref{figure1}D-E,iii,Fig. S4-5).
Interestingly, the ratio of these length scales \(R^*_{\nu}/R_{\vec{S}}^*\) increases over time reaching its maximal value of \(\sim 0.9\) just prior to the onset of \(\langle\vec{\nabla}\cdot \vec{u}\rangle < 0\) (\cref{figure2}D, open blue squares). 
Prior work has shown in strongly driven F-actin networks that the rate of energy dissipation and filament bending energy increases until the onset of contractility \cite{seara_entropy_2018}. In this light, we surmise that the increase of \(R^*_{\nu}/R_{\vec{S}}^*\) reflects the decoupling of the pressure and shear deformation fields via filament bending until bulk contraction once again suppresses these fluctuations. 

\clearpage 
\subsection{Temporal Decorrelation of Responses Enumerate Modes of Dissipation Across Length Scales}
\begin{figure}
    \centering
    \includegraphics[width=0.75\textwidth]{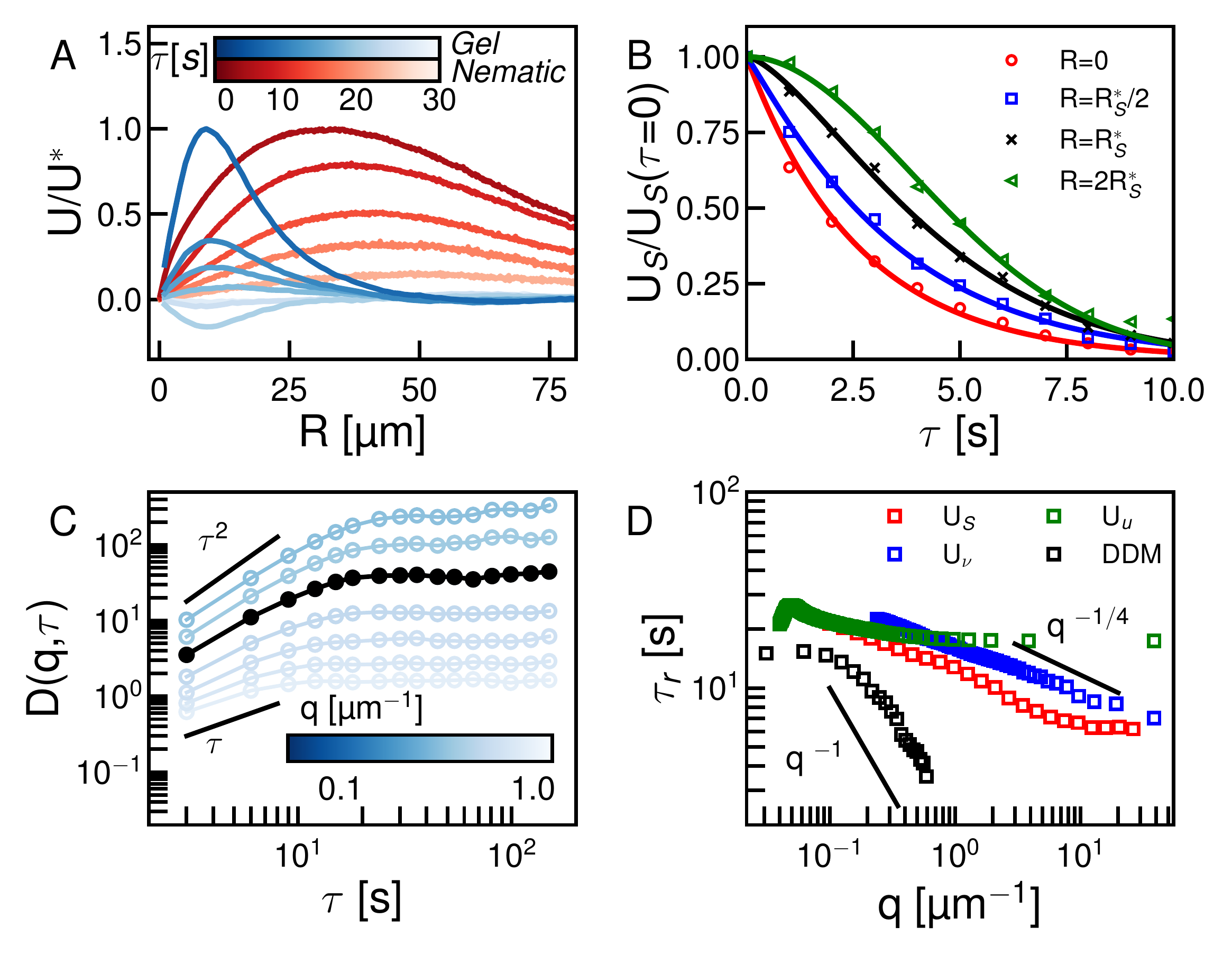}
    \caption{\textbf{Temporal dependence of correlated displacement field reveals characteristic time scales of active materials}. A) Normalized one dimensional shear response, \(U_{\vec{S}}\), for an active nematic (blue) and active gel (red) measured at different delay time \(\tau\) (see scale bar). B) Normalized shear response as a function of delay time \(\tau\) measured at different spatial scales with respect to the critical length scale \(R^*\). Experimental data is indicated with symbols solid lines are fit of the data to a stretched exponential. C) Dynamic structure function \(D(q,\tau)\) at different wave number \(q\) (see scale). \(D(q=2\pi/R^*,\tau)\) is plotted in black. \(\tau^2\) and \(\tau\) scaling are indicated. D) Characteristic time scales \(\tau_r\) as a function of wave number \(q\) for DDM (black) and displacement (\(U_{u}\), green), shear (\(U_{S}\), red), or vortical (\(U_{\nu}\), blue) response. 
    }
    \label{figure3}
\end{figure}

We next investigated dissipation across different length scales.
Recall that in the active nematic, all the curves of \(U_{\vec{S}}(\tau)\) as a function of lag time (\cref{figure1}E,iv \cref{figure3}A, red curve), shared a similar form. Namely, \(U_{\vec{S}}^*\) decreases and \(R_{\vec{S}}^*\) increases as \(\tau\) grows, until at longer lag times the response is negligible (\cref{figure3}A, red curves; lighter colors indicate longer lag times). This steady decorrelation in time is characteristic of viscous dissipation. 
In contrast to this monotonic decay, \(U_{\vec{S}}^*\) in the contractile gel becomes \textit{negative} at intermediate lag times before approaching zero (\cref{figure3}A, blue curves). This anticorrelated signal is a strong signature of elasticity with the reduced peak height reflecting contributions from viscous dissipation. This contractile gel then bears a strong signature of viscoelasticity dissipation; biochemical modifications to the contractile gel that enhance viscous relaxation abrogate this response (Fig. S5). 

We quantify the length-scale-dependent dissipation in the active nematic by considering the normalized relaxation of \(U_{\vec{S}}(\tau)\) at several distances \(R\) with respect to the peak \(R_{\vec{S}}^*\) (\cref{figure3}B).
Each relaxation is fit to a scaled exponential function, \(U(\tau)=\exp(-[\tau/\tau_r(R)]^{\gamma(R)})\); where \(\tau_r(R)\) is the size dependent decay time and \(\gamma(R)\) is the size dependent scaling factor. We define the decay time from scale-dependent response measurements, \(\tau_r(q=2\pi/R)\) as the time that the normalized response profile (\cref{figure3}B) relaxes to the inverse of the Euler number, \(\sim 0.368\).
For \(R = 0\) we find that the scaling factor \(\gamma=1\) consistent with diffusive relaxation of the shear response (\cref{figure3}B, red curve). At large scales, \(R = 2R_{\vec{S}}^*\), the relaxation profile is a stretched exponential \(\gamma=2\), diagnostic of ballistic type motion \cite{cipelletti_universal_2000, lee_myosin-driven_2021} (\cref{figure3}B, green). At intermediate scales, \(R_{\vec{S}}^*/2 \le R \le R_{\vec{S}}^*\), \(\gamma\) transitions smoothly between these extremes, indicating the transition between stochastic and flow dominated dissipation in the active nematic (\cref{figure3}B, blue and black).

To put these results in context, we compare our results to measurements from the established technique of differential dynamic microscopy (DDM) \cite{cerbino_differential_2008,reufer_differential_2012,lee_myosin-driven_2021}. We calculate the dynamic structure function \(D(q,\tau)\) from time lapse images of the active nematic considered above over a range of wave-numbers \(q\), from \(\SI{0.1}{\per\micro\meter}\) to \(\SI{1}{\per\micro\meter}\) and lag times \(1 < \tau < \SI{200}{\second}\) (\cref{figure3}C, color indicates \(q\)). We define a characteristic timescale \(\tau_r(q)\) as the lag time when \(D(q,\tau)\) reaches \(63.2\%\) of  \(D(q,\infty)\); where \(D(q,\infty)\) is the plateau value at long lag time. For short lag times, \(\tau < \tau_r(q)\), \(D(q,\tau)\) follows a power law, \(\tau^{\gamma_{DDM}}\) reflecting high frequency relaxation. Consistent with our response function analysis, we find \(\gamma_{DDM} = 2\) for \(q < \SI{0.1}{\per\micro\meter}\), and \(\gamma_{DDM} = 1\) for \(q > \SI{1}{\per\micro\meter}\)  indicating ballistic type motion at long length scales and diffusive type motion at short length scales (\cref{figure3}C). Similar to scale dependent response functions, we find a smooth transition between these behaviors in intermediate regimes (\cref{figure3}C).

While both methods agree on qualitative scaling at extreme scales we find that the critical time scales \(\tau_r(q)\) extracted from DDM are uniformly shorter than relaxation times found using our response analysis (\cref{figure3}D, black squares DDM, colored squares response analysis). Furthermore, this is not the product of a mere baseline shift but rather of markedly different scaling (\cref{figure3}D). For large \(q\) the relaxation time measured by DDM scales as \(\tau\sim 1/q\). Over the same range \(U_{\nu}\) becomes q independent (\cref{figure3}D, open green squares). This disconnect can be understood as DDM relaxation necessarily combines advection of material structure and dynamic relaxation processes. As such DDM correlations relax quickly. In contrast, \(U_{\nu}\) stops scaling below the agent size in the system --  \(\sim \SI{1}{\micro\meter}\) here (\cref{figure3}D, green) -- while \(\tau_r\) from shear and vortical responses scale weakly with wave number in small scales, \(\tau_r\sim q^{-0.25}\) (\cref{figure3}D, blue and red). This slower scaling of \(\tau_r(q)\) reflects the relaxation of only one source of strain. While DDM combines many modes of relaxation, response functions allow us to tease apart relaxation from many sources of stress independently.

\clearpage 
\subsection{Response functions differentiate modes of cellular contractility}
\begin{figure}
    \centering
    \includegraphics[width=0.75\textwidth]{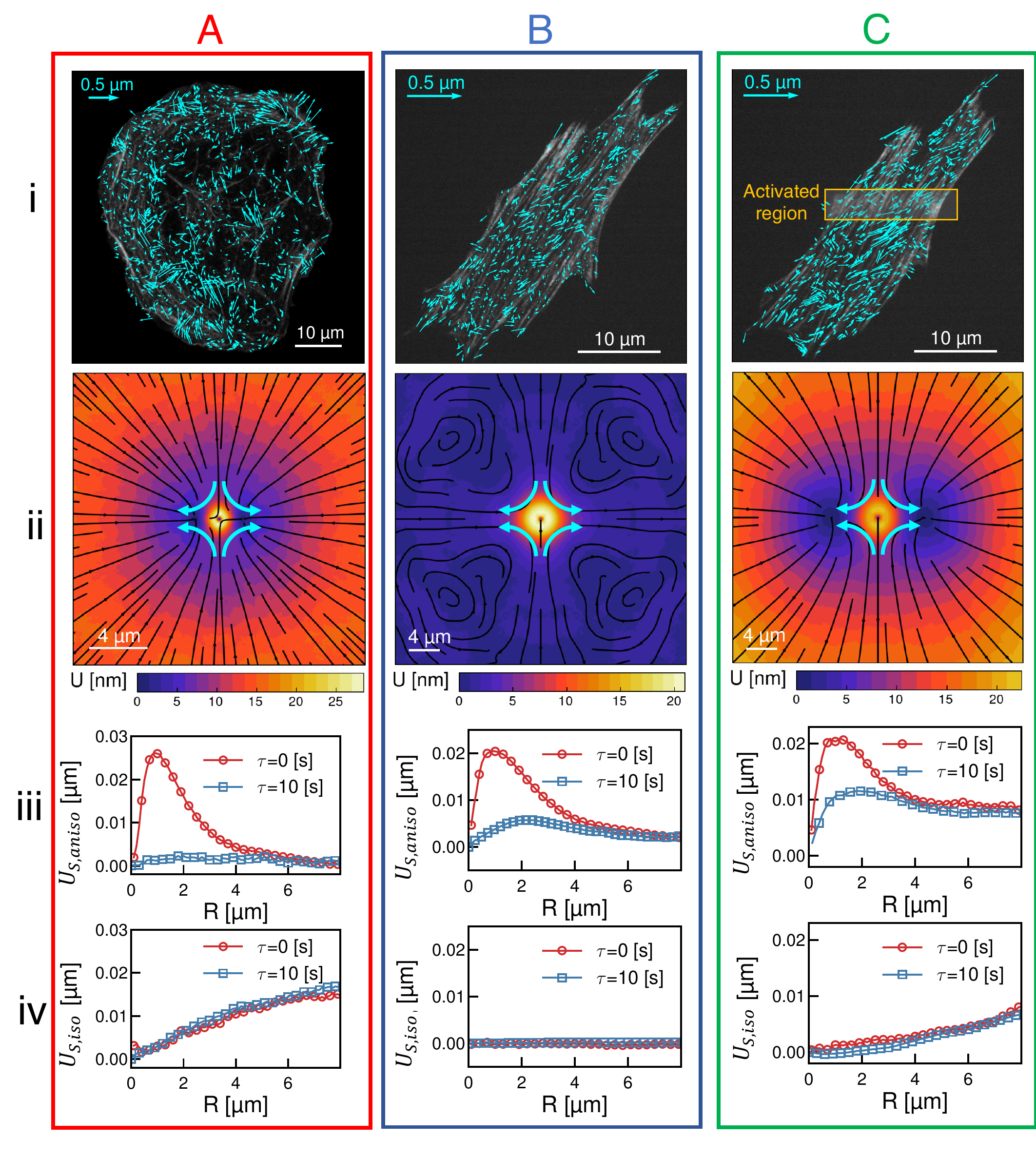}
    \caption{\textbf{Utilizing response functions to differentiate mechanical response of the actomyosin cytoskeleton in living cells.}  
    (i) Micrographs of the actin fibers (grey) overlaid with myosin displacement vectors (blue) for A) transverse arcs (in a U2OS cell), B) ventral stress fibers (in an NIH 3T3), and C) regional stimulated ventral stress fibers (NIH 3T3). The orange box in (C,i) indicates the region of optogenetic activation. 
    (ii) Shear response \(\vec{U}_{\vec{S}}\) for the displacement fields shown in (i). Streamlines indicate the direction of the response and color indicates the magnitude (see color bars). 
    (iii, iv) One dimensional traces of the anisotropic, \(\vec{U}_{\vec{S}, aniso}\) (iii), and isotropic \(\vec{U}_{\vec{S}, iso}\) (iv), parts of the shear response shown in (ii) measured at lag times of \(\tau=\SI{0}{\second}\) (red) and \(\tau=\SI{10}{\second}\) (blue).}
    \label{figure4}
\end{figure}

We now explore whether our method can distinguish different modes of actomyosin contractility in living cells. In adherent cells, the actomyosin cytoskeleton is organized into networks and bundles with highly stereotyped architecture and dynamics \cite{tojkander_actin_2012}. Transverse arcs are actomyosin bundles formed near the cell periphery and oriented parallel to the cell edge. Myosin activity continually drives the coalescence and contraction of transverse arcs resulting in their in continual inward motion, a dynamic process known as retrograde flow \cite{hotulainen_stress_2006,tojkander_actin_2012}. In contrast, ventral stress fibers are highly stable actomyosin bundles anchored on each end by focal adhesions  \cite{hotulainen_stress_2006,tojkander_actin_2012}. While both architectures can coexist, broadly circular U2OS osteosarcoma cells and elongated NIH 3T3 fibroblast cells predominantly display transverse arcs and ventral stress fibers, respectively (\cref{figure4}A,B). In each cell type, timelapse imaging of fluorescently labeled myosin is used to obtain displacement fields of the actomyosin dynamics (\cref{figure4}A,B, stress fibers grey, flow field blue arrows). 

The shear response \(\vec{U}_{\vec{S}}\) is then calculated for different lag times (\cref{figure4}ii) and as in \cref{figure2}E is split into one dimensional traces of the anisotropic \(U_{\vec{S}, ansio}\) (\cref{figure4}iii) and isotropic \(U_{\vec{S}, iso}\) (\cref{figure4}iv) components. The most dramatic difference between these architectures is in the isotropic shear response. In transverse arcs this response increases linearly as a function of distance and does not diminish over time, reflecting long range and coherent retrograde flow (\cref{figure4}A,ii,iv). The isotropic shear response is entirely absent in ventral stress fibers reflecting a lack of contractile dynamics (\cref{figure4}B,ii,iv). In contrast to the differences in the isotropic shear response, in both cases the anisotropic shear response at \(\tau = \SI{0}{\second}\) is peaked around \(\SI{1}{\micro\meter}\) (\cref{figure4}A-B,iii red). This reflects a similar length scale of maximal shear distortion. By \(\tau = \SI{10}{\second}\) the anisotropic shear response in transverse arcs decays completely (\cref{figure4}A,iii blue). This indicates that shear deformations in transverse arcs decorrelate faster than the time scale \(\tau\). In contrast, in ventral stress fibers \(U_{\vec{S}, ansio}\) persists at \(\tau = \SI{10}{\second}\) (\cref{figure4}B,iii blue) reflecting longer lived shear distortions. The disparity in anisotropic shear response at later lag time reveals differences in local dissipation of shear stress arising from differences in boundary conditions and local mechanical properties.

Both ventral stress fibers and transverse arcs reflect steady state dynamics common in adherent cells. Perturbations around these steady states can be queried via recently developed optogenetic techniques. Regional activation of RhoA in NIH 3T3 cells drives increased local actomyosin contractility and induces flow in adjacent ventral stress fibers \cite{oakes_optogenetic_2017}. Using this previously published data, we measure the shear response of ventral stress fibers prior to (\cref{figure4}B) and during (\cref{figure4}C) regional RhoA activation in the orange box indicated in \cref{figure4}C,i. Activation induces a shear response qualitatively similar to transverse arcs but with some important distinctions. The most notable similarity is that upon optogenetic activation, ventral stress fibers exhibit a linearly increasing isotropic shear response that is consistent over \(\tau = \SI{10}{\second}\) but with a decreased magnitude (\cref{figure4}C,iv). This underscores that a spatial gradient of contractile stress is sufficient to induce long range coherent flows in ventral stress fibers (\cref{figure4}C,iv). Interestingly, the anisotropic shear response at short scales of ventral stress fibers with and without regional activation are qualitatively similar (\cref{figure4}A-C,iii). In longer length scales, however, while the anisotropic response decays to 0 in the case without regional activation, the regional activation induces a far field response that does not fully decay reflecting the departure of the system from steady state dynamics (\cref{figure4}A-C,iii). This deviation from steady state is similar to what was observed in the case of the active gel as divergence increased (\cref{figure2}E, solid lines). The short and long range behavior of \(U_{\vec{S}, ansio}\) over lag times allow us to discern local mechanical properties and degree of mechanical homeostasis in cellular actomyosin architectures.

\clearpage 
\section{Conclusions}

Here we have introduced a method that extends correlation analysis to probe the full response function of active systems. Traditional correlation analysis takes into account only the position of each element of the field over which the correlation is measured, averaging over field elements that may point in different direction to produce radially symmetric correlation functions. Here we overcome this inherent dimensionality reduction by constructing correlations in which we align the correlation average with each field element in turn. By aligning our correlations, we take into account the response of the field in multiple dimensions and effectively measure a material response function from data. While we focused here on the response of a displacement field \(u\) to some derivative of itself, this method can be applied more generally. In fact, we can apply the same method to measure the response of any tensor field \(\vec{q}\) to a meaningful perturbative field \(\vec{p}\). We can thus extend the basic framework presented here to gain insights into other many body systems with complex dynamics such as high order quantum systems. 

To demonstrate the utility of this technique we compared the response of the displacement field in active actomyosin materials to previously established methods. We found that the critical length scale of the vorticity response in an active nematic, \(R_{\nu}^*\), is equal to the average vortex radius measured with traditional analyses \cite{giomi_geometry_2015}. We decomposed the shear response of an active gel into anisotropic and isotropic parts and found that the isotropic part captured the same contractions as the divergence of the velocity field \cite{murrell_f-actin_2012}. Furthermore, our method enumerated a transition from diffusive like relaxation at small scales to ballistic like relaxation at large length scales in agreement with previous analysis \cite{reufer_differential_2012}. In addition to these expected results we also found that the ratio of vortical to shear length scales presaged the onset of contractility in active gels. This provides a window into stress buildup and propagation and how it relates to the onset of compressibility in contractile gels. The temporal relaxation of shear responses allowed us to capture viscoelastic relaxation in multiple materials. This enables future work to address how the peculiarities of local driving affect long range dissipation in active system. Finally, we also demonstrated the utility of this method to measure material properties of the actomyosin cortex in vivo. This represents a powerful new tool to query cell mechanics. 

The specific measurements presented in this work are in many ways simply the lowest hanging fruit. A great deal of promise remains to be unlocked by combining the analysis presented here with specific experimental perturbations. By measuring response functions over a range of conditions one could calibrate the method and allow for a truly quantitative understanding of complex systems. In nematics alone one could imagine using such a procedure to detail the effect of substrate coupling, filament length, or even active agent composition on mechanics and dynamics. In cells, one could imagine genetically perturbing an actin accessory protein and specifically understanding how the mechanics of the network respond to that perturbation. Because the method isn't based on any specific physical model we can apply it to these systems and many more, including those with unknown mechanics. As such, we envision this method being of great use from dynamic systems to cell biology.

\clearpage

\section{Acknowledgements}
The authors thank Daniel Seara for insightful comments. This work was partially supported by the University of Chicago Materials Research Science and Engineering Center, which is funded by National Science Foundation under award number DMR-2011854. M.L.G. acknowledges support from NSF DMR-1905675,DMR 2215605, OMA-2121044 and NIH R01GM104032. J.J.D.P. acknowledges support from NSF Grant DMR-1710318. S.R. was supported by NIH T32 EB009412. This work was completed in part with resources provided by the University of Chicago Research Computing Center.

\section{Author Contributions}

M.M. developed the method; M.M., S.A.R., and M.L.G. analyzed the data; S.A.R., W.C., D.S., and P.O. performed experiments; S.A.R, M.M., W.C, and M.L.G. wrote the manuscript.

\section{Conflicts of Interest}
The authors declare no conflicts of interest.

\section{Materials and Methods}

\subsection{Active nematics and gel preparation}

Nematic experiments were performed as described previously \cite{zhang_spatiotemporal_2021}. Gel data in figure 3A is from \cite{murrell_f-actin_2012}. Briefly, \(\SI{2}{\micro\meter}\) actin 10\% labelled with tetramethylrhodamine-6-maleimide (TMR) was polymerized in F-buffer [10 mM imidazole, 1 mM MgCl2, 50 mM KCl, 0.2 mM egtazic acid (EGTA), pH 7.5] in the presence of 1mM  (gel) or \(\SI{100}{\micro M}\) (nematic) ATP. To minimize photobleaching, an oxygen scavenging system (4.5 mg/mL glucose, 2.7 mg/mL glucose oxidase(cat$\#$345486, Calbiochem, Billerica, MA), 17000 units/mL catalase (cat $\#$02071, Sigma, St. Louis, MO) and 0.5 vol. $\%$ $\beta$-mercaptaethanol is added to the actin mixture. 0.3\% w\% 400 cP methylcellulose is added to this mixture to crowd actin filaments to the bottom of the sample volume. Nematic samples also included 30nM f-actin capping protein to limit filament growth while gels were uncapped. Nematic samples were driven by 100nM synthetic tetrameric motors as described in \cite{schindler_engineering_2014}, while gels were driven by 50nM rabbit skeletal muscle Myosin II \cite{murrell_f-actin_2012}.

The sample was imaged on an Eclipse-Ti inverted microscope (Nikon, Melville, NY) in confocal mode utilizing a spinning disk (CSU-X, Yokagawa Electric, Musashino, Tokyo, Japan) and a CMOS camera (Zyla-4.2 USB 3; Andor, Belfast, UK). Nematic experiments were imaged collecting one frame every 2 seconds, while gels were imaged one frame every 5 seconds. 

\subsection{Flow field measurement}
In nematics velocity fields were calculated using the method of optical flow detailed in \cite{sun_secrets_2010} using the Matlab code available at (https:// ps.is.mpg.de/code/secrets-of-optical-flow-code-for-various-methods) and the ‘classic+nl-fast’ method. These predictions were processed into proper units in Matlab. Velocity fields in cells were determined using Quantitative Fluorescence Speckle Microscopy, QFSM \cite{ji_tracking_2005, danuser_quantitative_2006}.

\subsection{Response function measurement}
The code utilized in this manuscript (and sample data) are all found at \href{https://github.com/Gardel-lab/ResponseFunction.git}{https://github.com/Gardel-lab/ResponseFunction.git}. In brief, gradients of the velocity fields \(\vec{\nabla}\vec{u}\) are calculated using localized polynomial fitting and finite element method for the fields measured by the optical flow and QFSM respectively. The eigenvalues, \(\lambda\), of the strain rate tensor, \(\vec{S}_{ij}\) which is symmetric part of the \(\vec{\nabla}\vec{u}\), is calculated  by solving the characteristic equation \(|\vec{S}-\lambda \vec{I}|=0\), and eigendirections are obtained by plugging \(\lambda\) in the systems of equation \(\vec{S}-\lambda \vec{I}=0\) where \(\vec{I}\) is the identity tensor. For the active nematic the eigendirection associated with the positive eigenvalue and for active gel and cell data the eigendirection associated with the negative eigenvalue are chosen as direction to setup the coordinate system for ensemble averaging as described in \cite{molaei_interfacial_2021}. The data measured close to the edge of the field of view is discarded to prevent error from optical flow measurement to propagate to the measured response functions. \(\vec{\nabla}\vec{u}\) and its decomposed components, \(\vec{S}\), \(\vec{\Omega}\), and \(\vec{D}\)  are measured over grid points with spacing large enough to prevent oversampling the data. The correlation length of these components in different systems are used to select the spacing; for example, we chose the distance where the normalized auto-correlation functions of vorticity or strain rate tensor drops to 0.5. In general, in the chaotic and noisy system the grid spacing should be smaller than grid spacing in highly coherent systems.

\subsection{Cell culture}
U2OS cells with NMIIA endogenously tagged with eGFP is a generous gift from Dr. Jordan Beach (Loyola University Chicago). U2OS cells were cultured in McCoy’s 5A Medium (Sigma-Aldrich) supplemented with \(10\%\) FBS (Corning) and \(\SI{2}{m M}\) L-glutamine (Invitrogen). NIH 3T3 (ATCC) were cultured in DMEM (Mediatech) supplemented with \(10\%\) FBS (Corning) and \(\SI{2}{m M}\) L-glutamine (Invitrogen). Myosin was visualized by transfecting an mApple-RLC (gift of Mike Davidson, University of Florida) plasmid. 

\subsection{Microscopy and live cell imaging}
For the imaging of myosin dynamics in U2OS cells, Airyscan imaging was performed on a Zeiss LSM 980 microscope equipped with the Airyscan 2 detector. Images were acquired using the MPLX SR-4X mode and processed by Zen Blue 3.0 software using the Airyscan processing feature with default settings. During live cell imaging, cells were mounted on an imaging chamber (Chamlide) and maintained at \(\SI{37}{\degree C}\). For live cell imaging, cell medium was replaced with Dulbecco’s Modified Eagle Medium (DMEM) without phenol red (Corning) supplemented with \(10\%\) FBS, \(\SI{2}{m M}\) L-glutamine, \(1\%\) penicillin-streptomycin, \(\SI{10}{m M}\) HEPES and \(\SI{30}{\micro\liter\per m\liter}\) Oxyrase (Oxyrase Inc.).

The data in figure 4B,C appears in \cite{oakes_optogenetic_2017}, the NIH 3T3 cell was imaged on an inverted Nikon Ti-E microscope (Nikon, Melville, NY) with a Yokogawa CSU-X confocal scanhead (Yokogawa Electric, Tokyo, Japan) and laser merge module containing 491, 561 and 642 nm laser lines (Spectral Applied Research, Ontario, Canada). Images were collected on Zyla 4.2 sCMOS Camera (Andor, Belfast, UK). Local recruitment using the optogenetic probe \cite{oakes_optogenetic_2017} was performed using a 405 nm laser coupled to a Mosaic digital micromirror device (Andor). Images were collected using a 60X 1.49 NA ApoTIRF oil immersion objective (Nikon). All hardware was controlled using the MetaMorph Automation and Image Analysis Software (Molecular Devices, Sunnyvale, CA). 

\clearpage

\bibliographystyle{unsrt.bst} 
\bibliography{Gardel_ResponseFunction.bib}

\newcommand{\SItext}{%
  \section*{Supporting Information Text}
  \stepcounter{SItext}
}
\renewcommand{\thefigure}{S\arabic{figure}}
\renewcommand{\theequation}{S\arabic{equation}}
\setcounter{figure}{0} 
\setcounter{equation}{0}
\clearpage
\section{Supplemental Text}
\subsection{General cross correlation function}

In the main text we consider a simplified correlation function. Here let us consider a general version of the spatiotemporal correlation function
\begin{equation}
\label{genral_corr}
    \vec{C}_{\vec{p}\vec{q}}(\vec{R},\tau)=\langle \vec{p}(\vec{r}_1,t) \vec{q}(\vec{r}_2,t+\tau) \delta(\vec{R}-\vec{r}_{12})\rangle_{\vec{r}_1,\vec{r}_2,t},
\end{equation}
here \(\vec{p}\) and \(\vec{q}\) denote dynamical fields measured at time \(t\) and \(t+\tau\) respectively. The relationship being queried is the tensor product between \(\vec{p}\) and \(\vec{q}\) at positions \(\vec{r}_1\) and \(\vec{r}_2\). \(\vec{R}\) is a location with respect to a coordinate system with its origin at \(\vec{r}_1\). In order to make this comparison meaningful for the system at large, we average over all positions \(\vec{r}_2\) for each origin \(\vec{r}_1\) and then take the average over observation time denoted as \(\langle \rangle_{\vec{r}_1,\vec{r}_2,t}\). For simplicity we will omit subscripts and simply denote this operation as \(\langle\rangle\). The role of \(\delta\) is to bin the average into proper ensembles based on the spatial location \(\vec{R}\). \(\delta\) is a finite width delta function and sets the spatial precision of the measurement. In an ideal case this function has infinitesimal width and, \(\delta\) is Kronecker delta function. \(\vec{p}\) and \(\vec{q}\) can either denote to same dynamic variable, as in the case of auto-correlation, or they can denote different quantities as in the case of cross-correlation. In Euclidean space, \(\vec{C}_{\vec{p}\vec{q}}(\vec{R},\tau)\) is basically the first mode of probability of observable \(\vec{q}\) at the space-time point of \(\vec{R}\) and \(\tau\) if we know the value of \(\vec{p}\) at the origin. Note that taking the tensor product of \(\vec{p}\) and \(\vec{q}\) contains all of the information that might be desired from a traditional correlation function with a different product. For example, the spatial velocity-velocity correlation function \(\vec{C}_{\vec{v}\cdot\vec{v}}(R)=\langle \vec{v}(\vec{r}_1)\cdot\vec{v}(\vec{r}_2) \delta(R-|\vec{r}_1-\vec{r}_2|)\rangle\)  is the trace of equal-time auto-correlation tensor \(\vec{C}_{\vec{v}\vec{v}}\) in which \(\vec{R}\) is converted to the scalar distance \(R\). The choice of product or the manipulation of the tensor product is one of the key choices that lend this correlation function its interpretability. The other crucial piece is the choice of the coordinate system for \(\vec{R}\). Note that, both of this choices depend on the physics of the system. In the rest of this section we give one example of simplifying high rank correlation tensor related to the systems discussed in the main text as a guideline for other problems. Note that, we are only addressing linear systems here to take advantage of their superposition property.

When \(\vec{p}\) is a tensor with a rank higher than zero, for example when it is a velocity vector or a stress tensor, choosing specific coordinate systems for forming conditional averaging is a crucial step. When \(\vec{p}\) is a vector field, the direction of the vector is an obvious choice as describe in the text. However, consider an example where \(p_{ij}\) is a second rank tensor and \(q_i\) is a vector; therefore, \(\vec{C}_{\vec{p}\vec{q}}\) is a third rank tensor in two dimensions. To fully resolve \(\vec{C}_{\vec{pq}}\) we need to construct 4 general correlation vector field of \(\vec{\chi}\). One can consider various decomposition of \(\vec{p}\) to form different meaningful \(\vec{\chi}\). For example if \(\vec{p}\) denotes a strain tensor its principle directions can be used to set up the coordinate system and its invariant can be used as scalar in the modified correlation functions. Seeking to understand how vortical, normal, and shear deformations propagate in the active system we choose to decompose \(\vec{p}=\vec{\nabla} \vec{u}\) into the anisotropic symmetric traceless strain rate tensor, \({S}_{ij}=(\partial_iu_j+\partial_ju_i)/2-\partial_ku_k I_{ij}/2\), the isotropic symmetric strain rate tensor \({D}_{ij}=\partial_ku_k I_{ij}/2\), and the circulation tensor \({\Omega}_{ij}=(\partial_iu_j-\partial_ju_i)/2\). Spatiotemporal correlation function for \(\vec{C}_{(\vec{\nabla u})\vec{u}}\) then can be written as   
\begin{equation}
\label{genral_corr_decomposed}
\begin{aligned}
    \vec{C}_{(\vec{\nabla u})\vec{q}}(\vec{R},\tau)
    &=\langle [\vec{S}(\vec{r}_1,t)+\vec{D}(\vec{r}_1,t)+\vec{\Omega}(\vec{r}_1,t)] \vec{q}(\vec{r}_2,t+\tau) \delta(\vec{R}-\vec{r}_{12})\rangle\\
    &=\langle \vec{S}(\vec{r}_1,t)\vec{q}(\vec{r}_2,t+\tau) \delta(\vec{R}-\vec{r}_{12})\rangle\\
    &+\langle \vec{D}(\vec{r}_1,t) \vec{q}(\vec{r}_2,t+\tau) \delta(\vec{R}-\vec{r}_{12})\rangle\\
    &+\langle \vec{\Omega}(\vec{r}_1,t) \vec{q}(\vec{r}_2,t+\tau) \delta(\vec{R}-\vec{r}_{12})\rangle. 
    \end{aligned}
\end{equation}
Where each term on the right hand side can be reduced to a vector field by the treatment described in the text. 
Note that in two dimensions \(\vec{C}_{(\vec{\nabla} \vec{u}) \vec{u}}\) has 8 degrees of freedom.  In an isotropic system with no odd deformation field eight degree of freedoms in \(\vec{C}_{(\nabla \vec{u}) \vec{u}}\) reduces to 4. In the active system described in the text, \(\vec{C}_{(\vec{\nabla} \vec{u}) \vec{u}}\) were decomposed to \(\vec{\chi}_{\vec{S}}\),  \(\vec{\chi}_{\nu}\), and \(\vec{\chi}_{D}\). Since \(\vec{S}\) is a symmetric tensor and traceless \(\vec{\chi}_{S}(\vec{R})\) is symmetric about both \(X\) and \(Y\) axes and has two degree of freedoms.  \(\vec{\chi}_{\nu}\) only has the angular component, and \(\vec{\chi}_{D}\) only has the radial component.  Therefore, the set of modified correlation functions \(\vec{\chi}_{\vec{S},~\nu ,~ D}\) fully captures the third rank correlated tensor field \(\vec{C}_{(\nabla \vec{u}) \vec{u}}\). 

\clearpage
\section{Supplemental Figures}

\begin{figure}[h]
    \centering
    \includegraphics[width=0.4\textwidth]{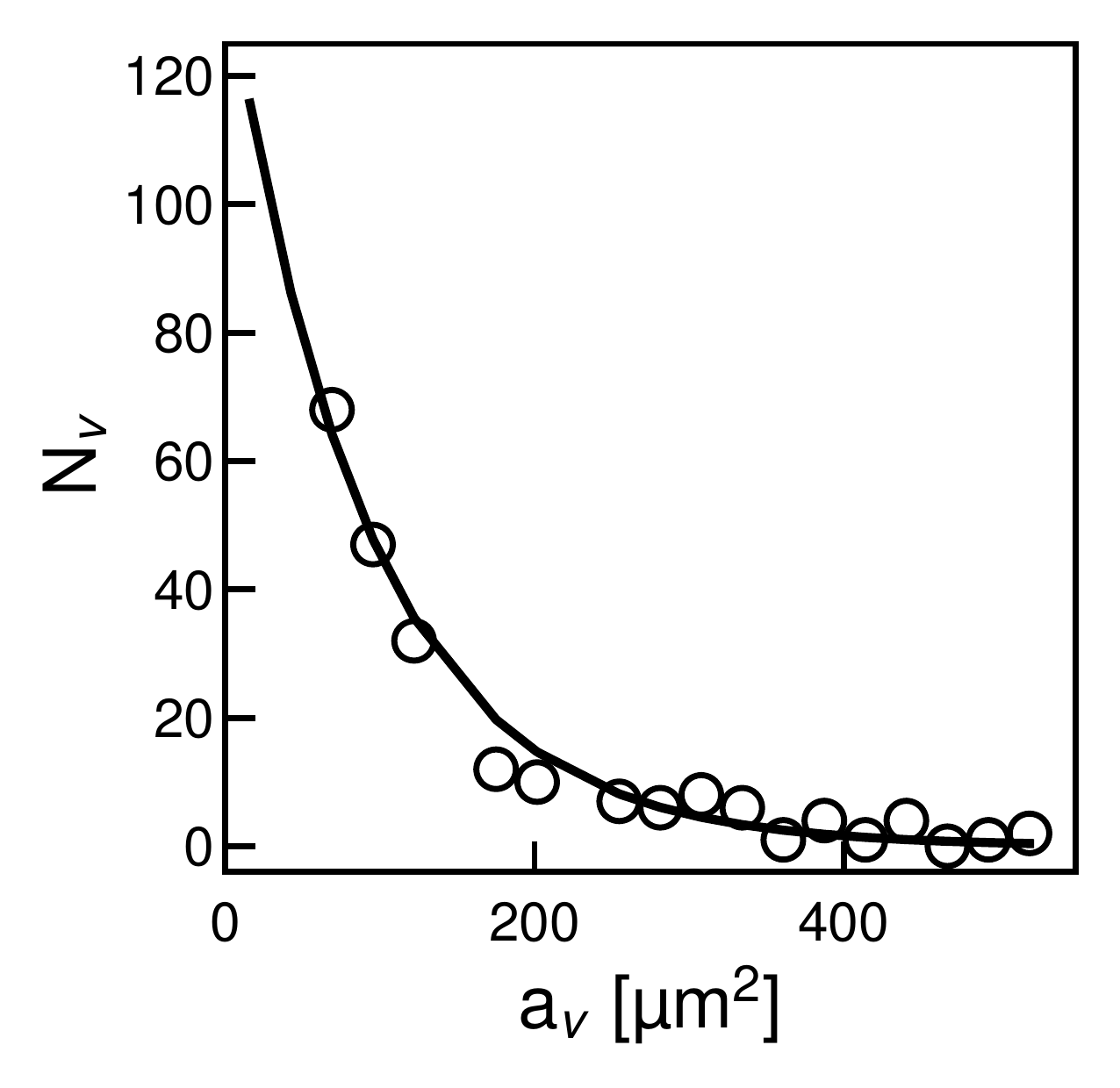}
    \caption{\textbf{Statistical analysis of vortex size}. Number of vortexes with different sizes, open circles, in the 2D active nematics of the short actin filaments studied in figure 1 in the main text. The solid line is an exponential fit to the data. The size of the field of view limits the range of vortex size measurement.}
    \label{VortexSize}
\end{figure}

\begin{figure}[h]
    \centering
    \includegraphics[width=0.7\textwidth]{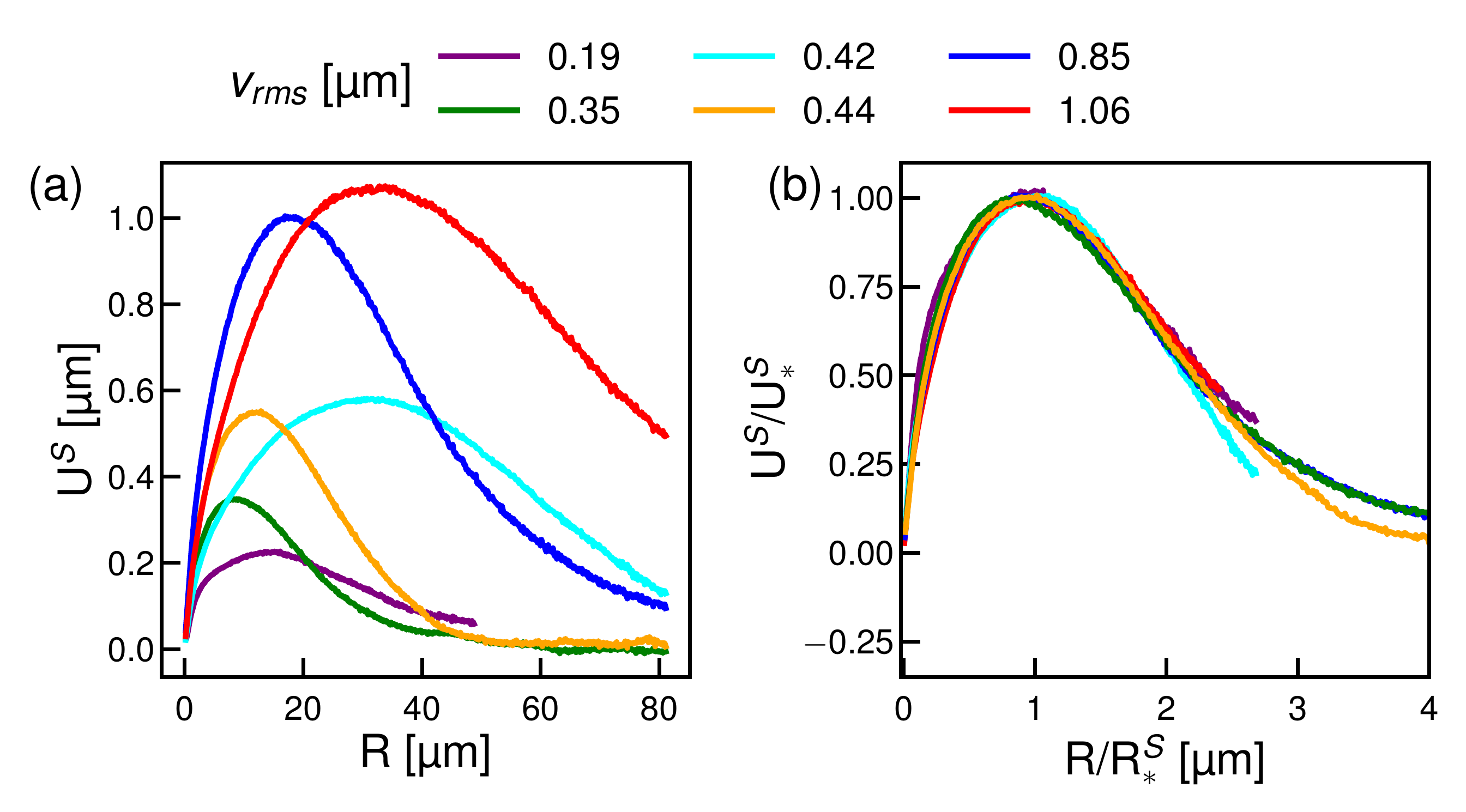}
    \caption{\textbf{Scaling behavior of active nematics:} Because critical scales are model-free and unambiguous they easily allow for comparisons between experimental conditions. a) Correlated deformation profiles of nematics with different rms velocity. b) Correlated deformation shown in (a) normalized with characteristic displacement scales \(U^S_*\) as a function of distance normalized with charactetic length scale \(R^S_*\).}
    \label{VortexSize}
\end{figure}

\begin{figure}
    \centering
    \includegraphics[width=0.4\textwidth]{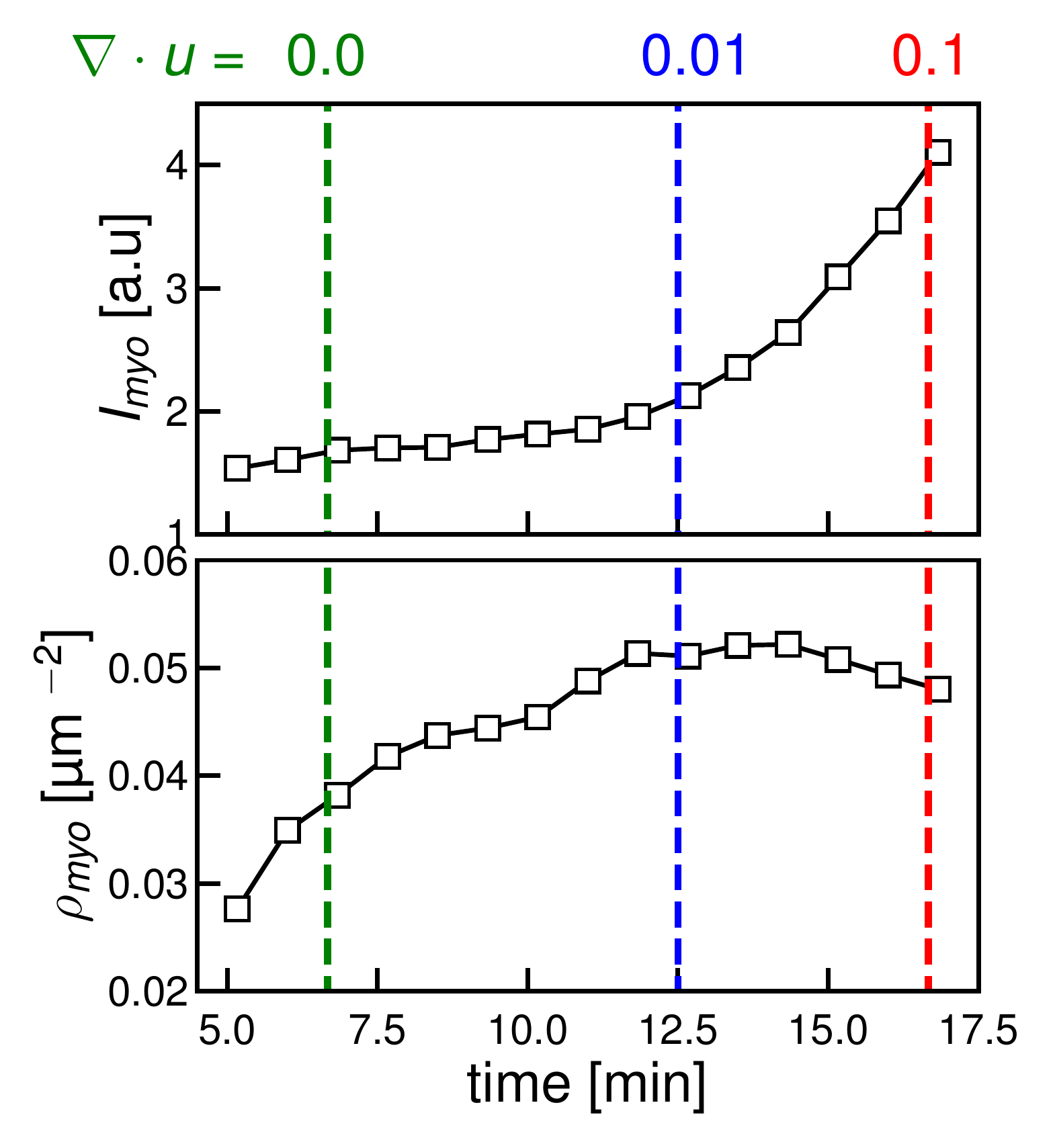}
    \caption{\textbf{Myosin puncta intensity and density in the active gel studied in figure 2 in the main text.} a) Optical intensity of the individual puncta proportional to their sizes. b) density of the myosin puncta over time of observation. The green, blue, and red dashed lines indicate the points taken as characteristic of the gel before contraction, at the onset of contraction, and deep in the contractile regime. }
    \label{MyosinDensity}
\end{figure}

\begin{figure}
    \centering
    \includegraphics[width=0.4\textwidth]{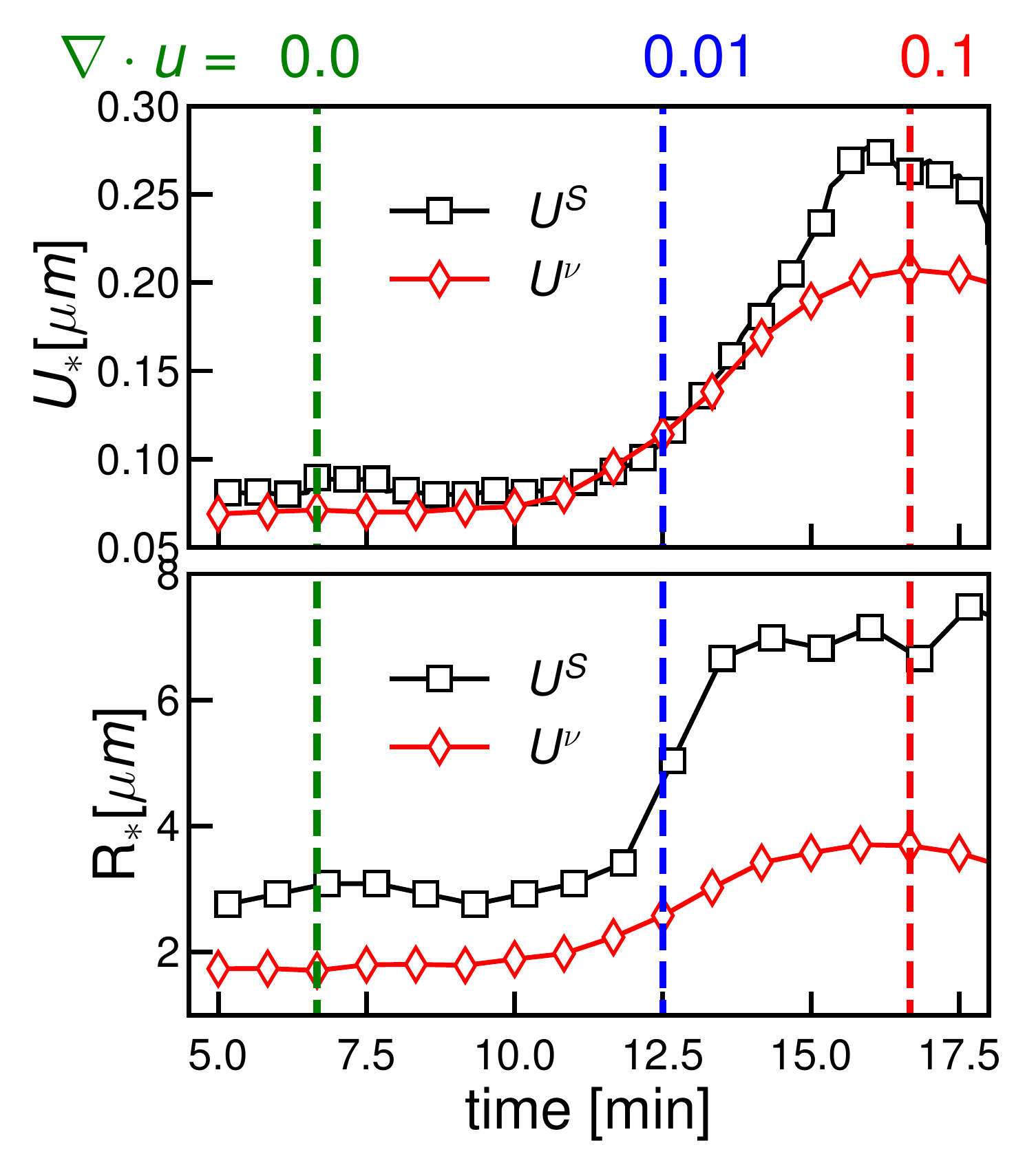}
    \caption{\textbf{Characteristic length and displacement scales of the active gel studied in figure 2 in the main text.} a) Characteristic displacement scales of the shear dipolar mode, \(U_*^S\) and vortical mode \(U_*^\nu\). b) Characteristic length scales shear dipolar mode, \(R_*^S\) and vortical mode \(R_*^\nu\) corresponding to the displacement scales of \(U_*^S\) and \(U_*^S\nu\). The green, blue, and red dashed lines indicate the points taken as characteristic of the gel before contraction, at the onset of contraction, and deep in the contractile regime. }
    \label{MyosinDensity}
\end{figure}

\begin{figure}
    \centering
    \includegraphics[width=0.4\textwidth]{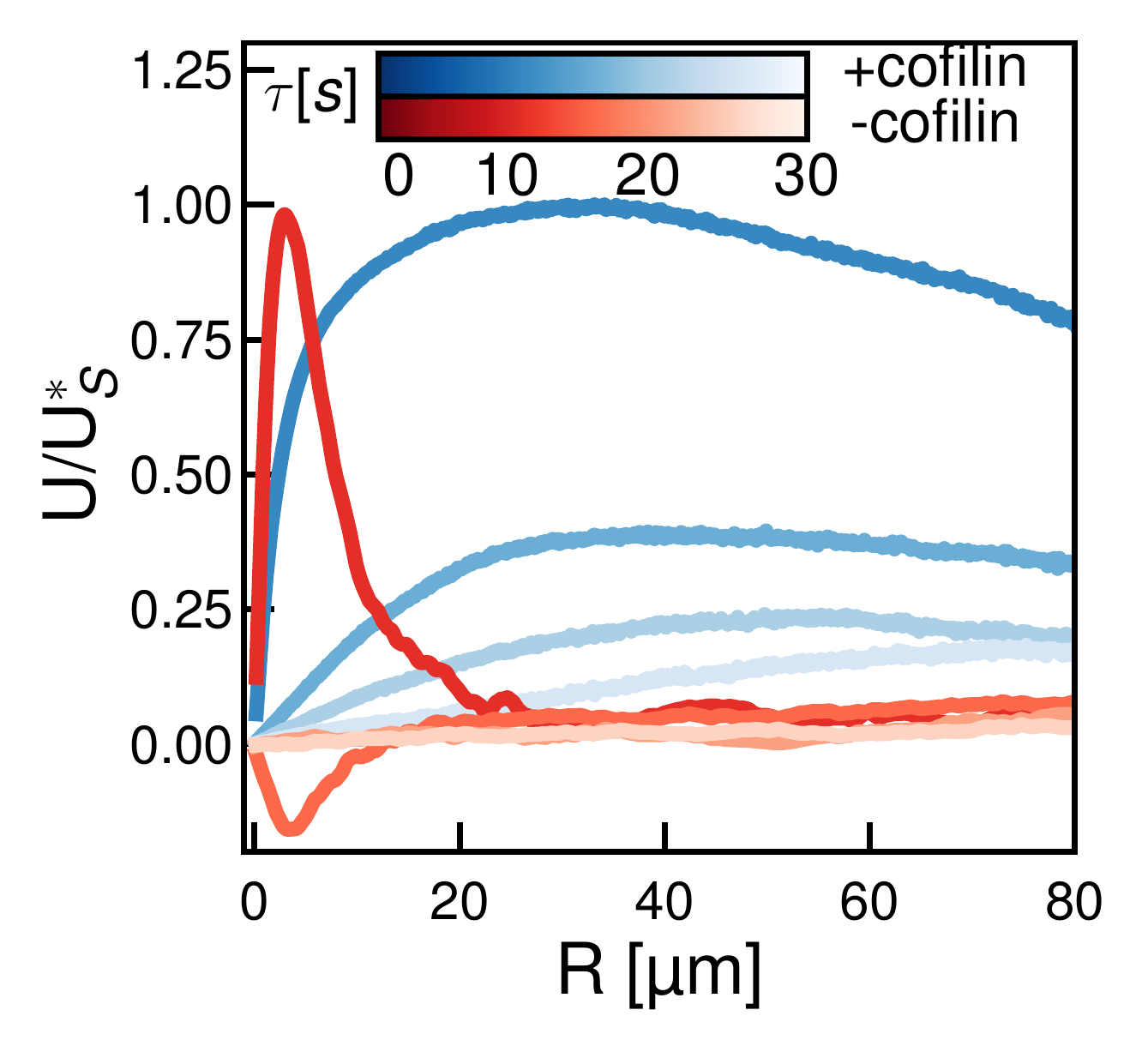}
    \caption{\textbf{Cofilin ablates the elastic response of an active gel} To confirm this interpretation, we compare gels with and without the addition of the actin severing protein cofilin which dissipates elastic stress through filament severing \cite{mccall_cofilin_2019}. We find that a gel with cofilin never anticorrelates as is consistent with its role in preventing stress buildup in actin networks. The one dimensional shear response \(U_{\vec{S}}\) as a function of lag time (lighter colors indicate later lag times) for an active gel with (blue) and without (red) the addition of the actin severing protein cofilin.}
    \label{cofilin}
\end{figure}

\begin{figure}
    \centering
    \includegraphics[width=0.9\textwidth]{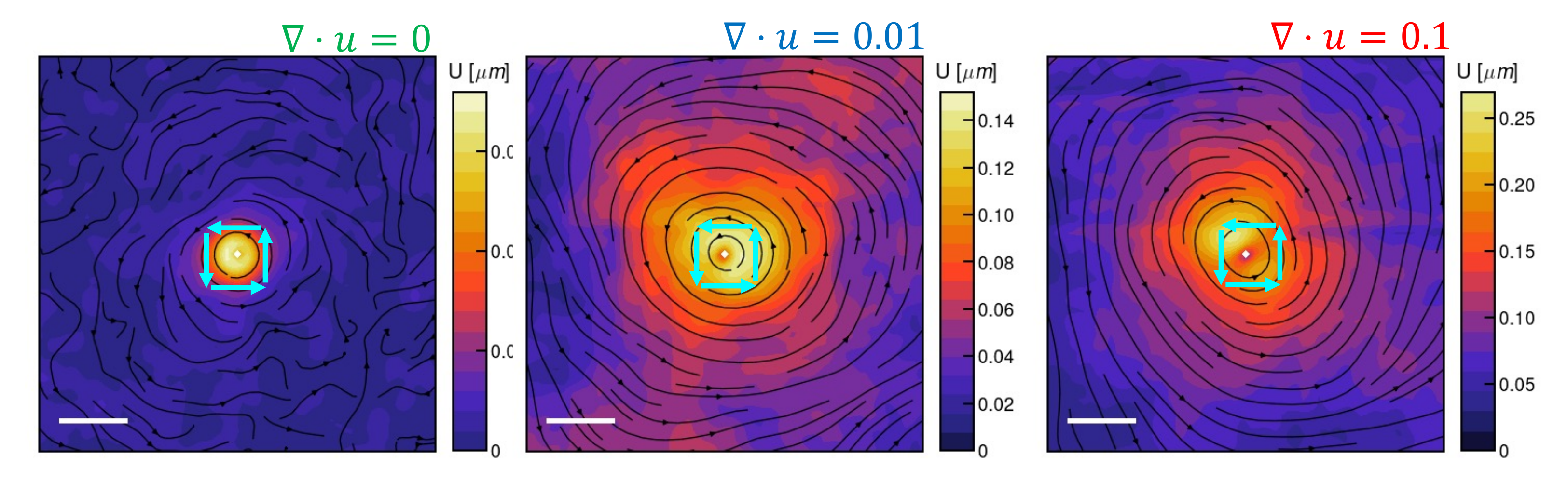}
    \caption{\textbf{Vortical deformation field of the active gel studied in figure 2 of the main text.} Vortical response of the active gel at the stable state with \(\nabla\cdot u=0.0\), at the onset of contraction with \(\nabla\cdot u=0.01\) and deep in the contractile regime with \(\nabla\cdot u=0.1\). Streamlines indicate the direction of the resulting correlation field and color indicates the magnitude; scale bars are \(\SI{5}{\micro\meter}\).}
    \label{MyosinDensity}
\end{figure}

\begin{figure}
    \centering
    \includegraphics[width=0.4\textwidth]{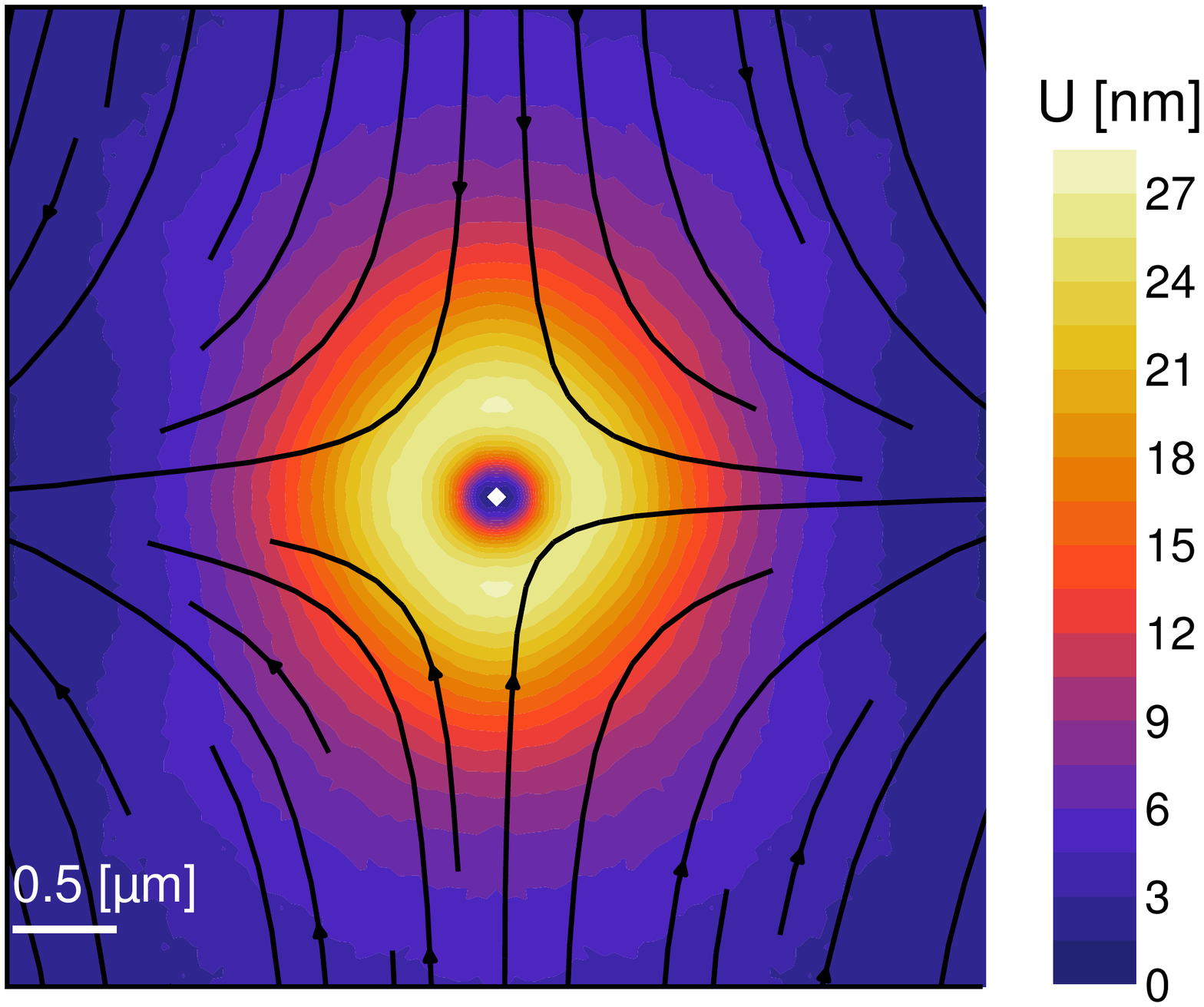}
    \caption{\textbf{Anisotropic shear dipolar deformation in U2OS cell}. Shear dipolar deformation field shown in figure 4C,ii after subtracting linear isotropic deformation field, revealing isolated local deformation field from the global retrograde flow.
    }
    \label{}
\end{figure}

\end{onehalfspacing}
\end{document}